\documentclass[11pt,dvips]{article}

\usepackage{epsfig,times}
\usepackage{picinpar}
\usepackage{wrapfig}
\usepackage{floatflt}
\makeatletter
\renewcommand{\section}{\@startsection{section}{1}{0in}
	{0.4\baselineskip}{0.1\baselineskip}{\Large\bf}}
\renewcommand{\subsection}{\@startsection{subsection}{2}{0in}
	{0.25\baselineskip}{-\baselineskip}{\large\bf}}
\renewcommand{\subsubsection}{\@startsection{subsubsection}{3}{0in}
	{0.1\baselineskip}{-\baselineskip}{\normalsize\bf}}
\makeatother
\newcommand{\apj}{ApJ}
\newcommand{\apjl}{ApJ}

\begin{document}

%
\thispagestyle{myheadings}
%
\markright{OG 2.4.9}
\begin{center}
%
{\LARGE \bf TeV Gamma Ray Emission from Cen X-3}
\end{center}

\begin{center}
%
%
{\bf P.M.~Chadwick, K.~Lyons, T.J.L.~McComb, K.J.~Orford,
J.L.~Osborne, S.M.~Rayner, S.E.~Shaw, and K.E.~Turver}\\
{\it Department of Physics, Rochester Building, Science Laboratories,
University of Durham, Durham, DH1~3LE, U.K.}
\end{center}

\begin{center}
{\large \bf Abstract\\}
\end{center}
\vspace{-0.5ex}
%
%
Cen X-3 is a well-studied high-mass accreting X-ray binary and a
variable source of high energy gamma rays from 100 MeV to 1 TeV. The
object has been extensively monitored with the University of Durham Mark
6 telescope.  Results of observations, including those taken in 1998 and 1999, are
reported. There is no evidence for time variability in all the VHE data.  There
is also no evidence for correlation of the VHE flux with the X-ray flux detected by BATSE and
{\em RXTE}/ASM. A search for periodic emission, at or close to
the X-ray spin period, in the VHE data yielded a 3$\sigma$ upper 
limit to the pulsed flux of $2.0 \times 10^{-12} {\rm~cm}^{-2} {\rm~s}^{-1}$.
%

\vspace{1ex}

%
%

\section{Introduction}

Results of observations of the accreting X-ray binary Cen X-3 using
ground based gamma ray telescopes have been reported which have included
evidence for sporadic outbursts of strong pulsed emission in the $>$ 1 TeV band
(Carraminana at al. 1989, Raubenheimer et al. 1989) and a constant but
weaker unpulsed emission at $> 400$ GeV (Chadwick et al. 1998). These
results, together with the {\it CGRO} EGRET measurement of an outburst
of pulsed GeV emission (Vestrand, Sreekumar \& Mori 1997), indicate
that Cen X-3, an accurately measured system containing a 4.8 s pulsar in
a 2.1 d orbit around an O-type supergiant, is a sporadic source of high
energy gamma rays. 

The discovery of very high energy (VHE) gamma rays from X-ray selected
BL Lacs has been a highlight of high energy astrophysics in recent years
(Weekes et al. 1997). These objects have been shown to be sporadic and
copious TeV gamma ray emitters. They exhibit extremely short term time
variability in TeV emission and a correlation between the emission of
X-rays and TeV gamma rays (Schubnell 1997). This short term TeV
variability and the correlation with X-ray emission provide constraints
on the possible production models for TeV gamma rays. Many
models involve the jets which are a feature of such objects. There have
been suggestions that some galactic objects may share the jet properties
more usually associated with AGNs (Hjellming \& Nan 1995) and jets have
been suggested as sources of TeV emission from X-ray binaries (Vestrand
\& Eichler 1982, Hillas 1984, Kiraly \& Meszaros 1988). 

We present the results of a search for a possible correlation between $>
400$ GeV gamma rays recorded by the University of Durham Mark 6
telescope and X-ray emission according to measurements made with the
{\it RXTE} and {\it CGRO}/BATSE experiments.  We report the results 
of analysis of data taken during 1998 March and April and 1999 February.  We also present the results
of searches for variation of the emission at both the orbital and spin periods.

\section{Recent Observations of VHE Gamma Rays}

We report observations made with the University of Durham Mark 6 imaging
gamma ray telescope operating at Narrabri NSW, Australia. The telescope
has been described in Armstrong et al. (1998) and the results of initial
observations of Cen X-3 have been reported (Chadwick et al. 1998). Our
Cen X-3 dataset now comprises data from 31 hrs of observation during
23 exposures in 1997 March and June (JD 2450508 -- JD 2450606), 1998
March and April (JD 2450899 -- JD 2450932) and 1999 February (JD 2451220
-- JD 2451230).

Our earlier report (Chadwick et al. 1998) was based on data recorded in
1997 March and June (JD 2450508 -- JD 2450606) only. Assuming a
collection area of $10^9 {\rm~cm}^2$ and that our selection procedure
retained $\sim 50\%$ of the original gamma ray events, the time averaged
flux was estimated to be $(2.0 \pm 0.3) \times 10^{-11} {\rm~cm}^{-2}
{\rm~s}^{-1}$ for $> 400$ GeV. We concluded that the measurements in these
two months were consistent with a constant flux. Ongoing simulations suggest
 that our current selection procedure retains $~ 20\%$ of the gamma
rays. On this basis, the flux for the 1997 March and June
(JD2450508 -- JD2450606) data would be $(5.0~\pm~0.9) \times 10^{-11}
{\rm~cm}^{-2} {\rm~s}^{-1}$. The additional data taken in 1998 and 1999
provide fewer gamma ray candidates suggesting weaker TeV emission. An
analysis of the total data yields a time averaged flux of
$(2.8~\pm~1.4_{sys}~\pm~0.6_{stat}) \times 10^{-11} {\rm~cm}^{-2}
{\rm~s}^{-1}$; the significance of the detection based on
the total dataset is $4.7 \sigma$.

\section{The TeV Gamma Ray Signal Strength}

\begin{wrapfigure}[19]{l}{8cm}  
\psfig{file=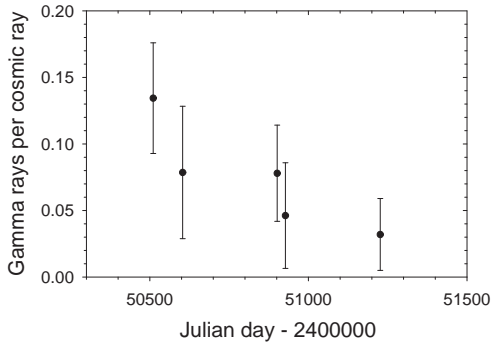,height=5.7cm}
\caption{The VHE gamma ray flux from Cen X-3 averaged over observing
periods.}
\label{fig:gammamonth} 
\end{wrapfigure}
Our recent work on PKS 2155--304 (Chadwick et al. 1999) has demonstrated
a method of assessing the signal strength of gamma rays recorded by
Cerenkov telescopes. It is suited to measurements made at different
epochs and at different zenith angles when the telescope may have
different sensitivities and consequently a varying background cosmic ray
detection rate. We have estimated the signal strength of TeV gamma ray
emission by expressing it as a fraction of the cosmic ray background
remaining after image shape and orientation selection (Fegan 1997). In
so doing we make allowance for variations in sensitivity, in first
order, due to changes in efficiency of the telescope and variations in
telescope performance with zenith angle. It is also assumed that the
slopes of the gamma ray and cosmic ray spectra are similar. 

In the present study, the average gamma ray signal strength from Cen
X-3, expressed as a percentage of the cosmic ray background remaining
after shape and orientation selection is $(7.0 \pm 1.5)$\%. The most
straightforward, but not most powerful, test for constancy of emission
is to repeat this process for the data recorded in each of the 5 dark
periods as shown in Figure \ref{fig:gammamonth}. On the basis of this
test we find no internal evidence for monthly variability of the VHE
signal; the data treated this way are consistent with a constant signal
strength $(\chi^2 = 4.5$, 4 df).

\section{X-ray Data}

Cen X-3 is a strong but variable X-ray emitter. For example, the average
daily rates for X-rays detected with the {\it RXTE}/ASM during 1997 and
1998 range from 0 to 32 counts ${\rm s}^{-1}$; the data are variable on
a time scale of days. The daily average for the {\it RXTE}/ASM count
rates are available for 22 of the 23 days when TeV gamma ray
observations were made\footnote{Available on the web at
\mbox{http://space.mit.edu/XTE/asmlc/srcs/cenx3.html}}.

The strength of pulsed X-ray emission was also available as a daily
average from the BATSE archive for 1997\footnote{Original data obtained
from the web at \mbox{http://www.batse.msfc.nasa.gov/data/pulsar}};
during the 1998 and some of the 1999 VHE observations, the X-ray flux was low and less than the threshold for
BATSE detection. The BATSE data provide a series of independent X-ray
measurements, including a measurement on the single day of the TeV gamma
ray observations for which there is no corresponding {\it RXTE}/ASM
measurement.

\newpage
\section{Correlations between X-rays and TeV Gamma Rays}

\begin{wrapfigure}[19]{l}{8cm} 
\epsfig{file=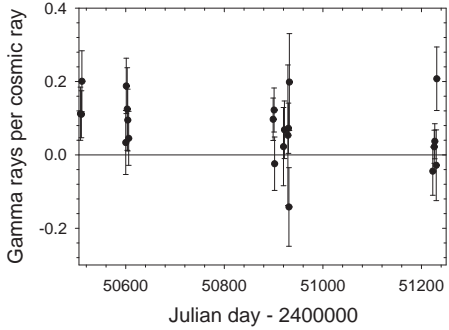,height=5.5cm}
\caption{The VHE gamma ray flux from Cen X-3 plotted on a day-by-day basis.}
\label{fig:gammaday} 
\end{wrapfigure}

The VHE gamma ray signal plotted on a day by day basis is shown in
Figure \ref{fig:gammaday}. There is no evidence for outbursts of TeV
gamma ray emission on a timescale of days and the data are consistent
with a constant TeV gamma ray flux ($\chi^2 = 22.1$, 22 df).

\begin{figure}[htb] 
\epsfig{file=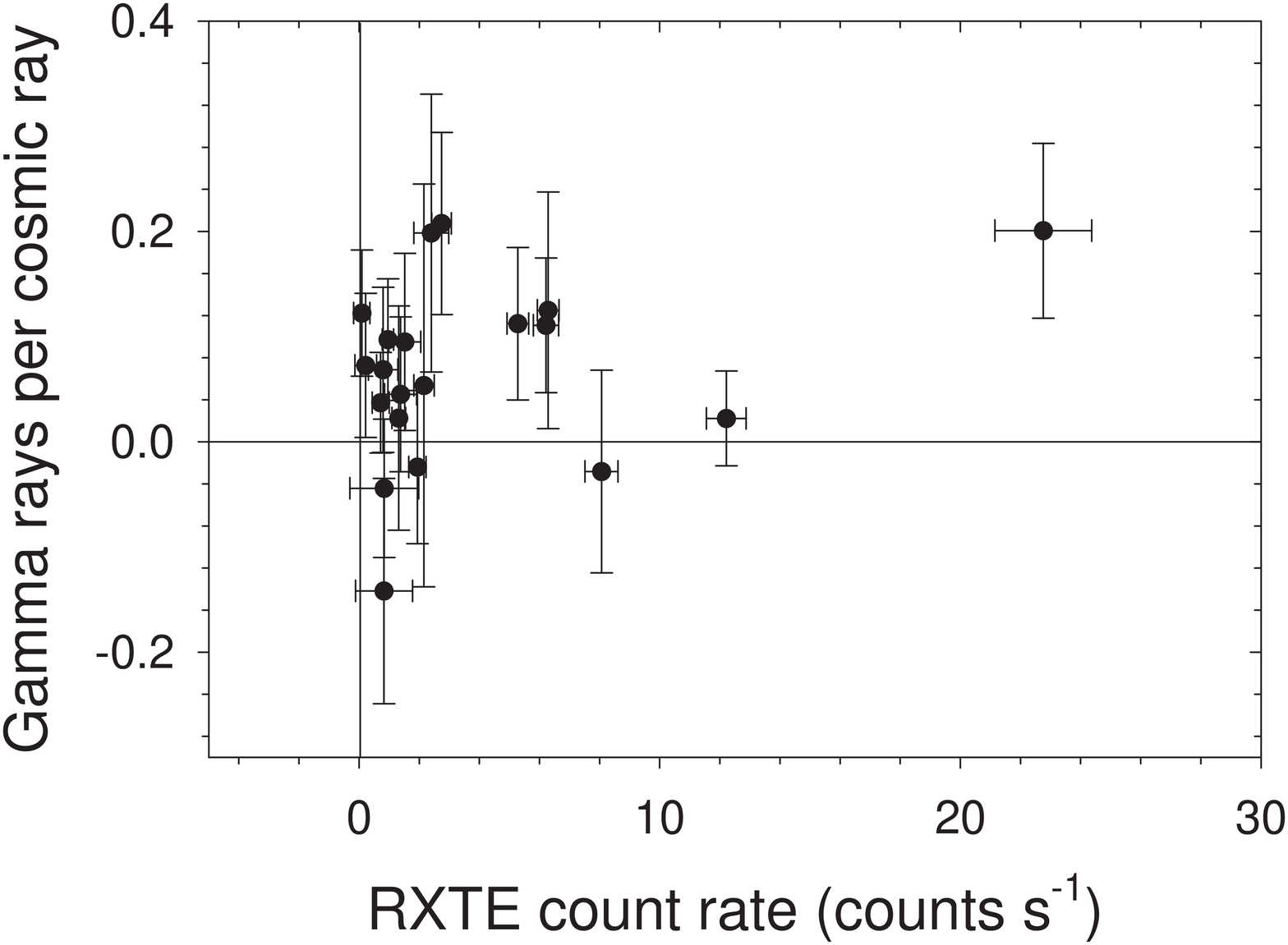,width=8cm}
\epsfig{file=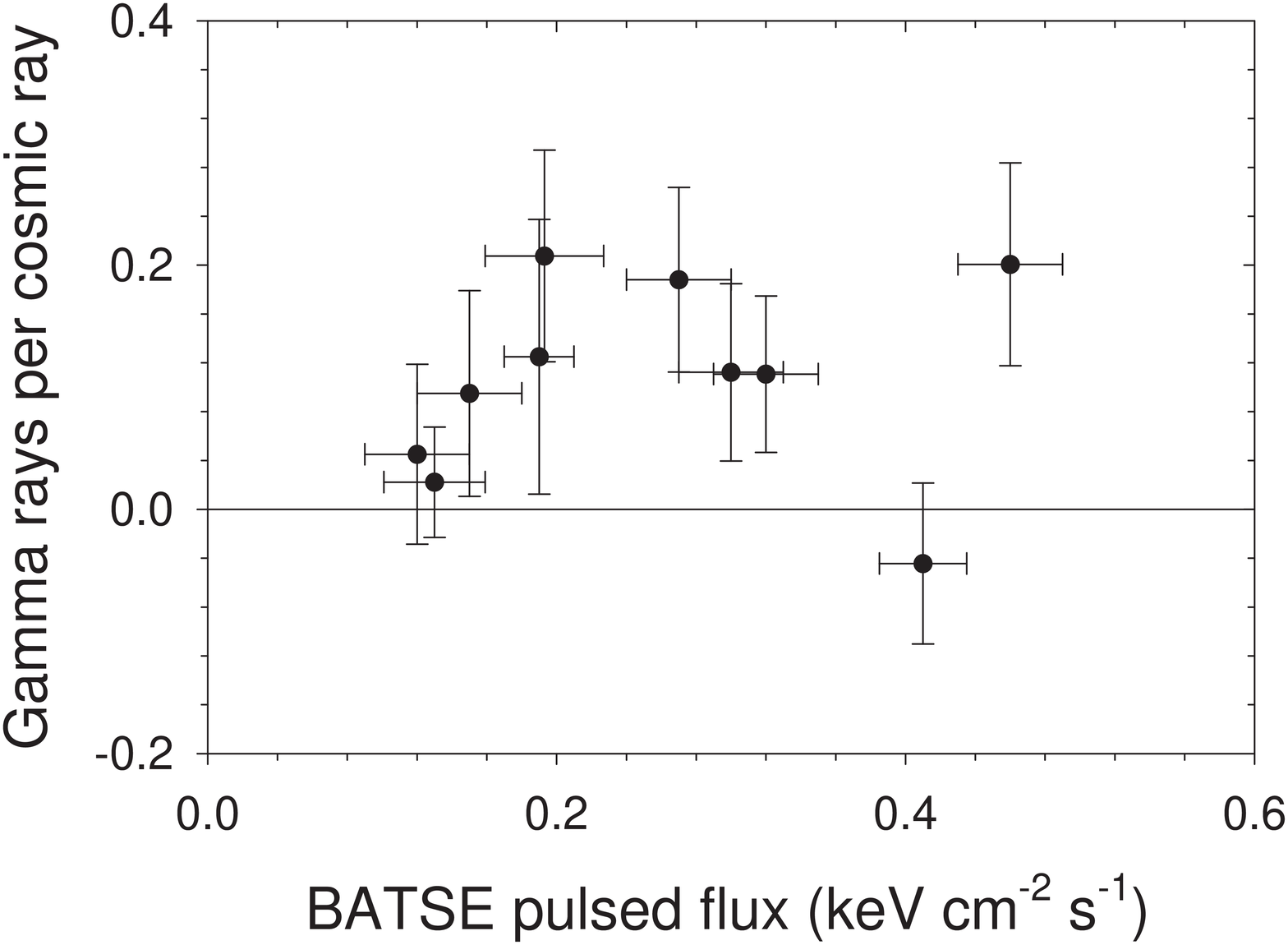,width=8cm}
\caption{The relation between the daily VHE gamma ray flux from Cen X-3
and (a) the X-ray flux detected by ASM/{\it RXTE\/} and (b) the X-ray pulsed flux detected by BATSE.}
\label{fig:daycor} 
\end{figure}

In Figure \ref{fig:daycor}(a) we show the relation between the count
rate of the {\it RXTE}/ASM data and our gamma ray signals. In Figure
\ref{fig:daycor}(b) we show a similar plot between the individual BATSE
pulsed X-ray fluxes and our gamma ray signals. We have no formal
evidence for a correlation, although it is interesting to note that the
day of highest detected gamma ray flux coincides with the day of most
X-ray activity in the dataset (1997 Mar 4).

\section{Modulation at the orbital period}

\begin{wrapfigure}[25]{l}{7.5cm} 
\epsfig{file=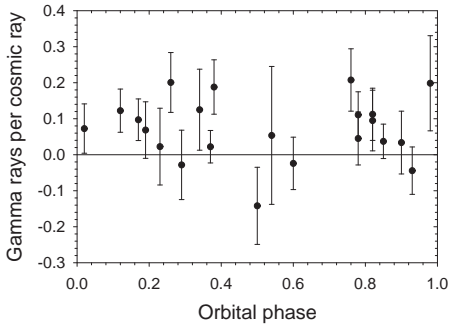,height=5.4cm}
\caption{The measured VHE gamma ray rate during each observation of Cen X-3 plotted as a function of orbital phase.}
\label{orbphase} 
\end{wrapfigure}

We have looked for modulation of the gamma ray signal at the orbital
period of the binary system. The orbital phase of each of our
observations has been calculated using the ephemeris of Kelley et al. (1983). The results
are shown in Figure \ref{orbphase}. From this evidence we conclude that
there is no modulation of the VHE gamma ray emission
at the orbital period.

\section{Modulation at the pulsar period}

The data have been subjected to a Rayleigh test for periodicity at a small
range of periods around the BATSE period.  Phase coherence 
between observations was not assumed.  No significant periodicity was detected, 
leading to a 3$\sigma$ upper limit to the pulsed flux of $2.0 \times 10^{-12} 
{\rm~cm}^{-2} {\rm~s}^{-1}$ in the total dataset.

\section{Discussion}

We have detected VHE gamma ray emission from Cen X-3 during each dark
moon period that we have observed this object. The data are consistent
with a weak but persistent emission, both when the VHE data is averaged
over dark moon periods or when considered observation by observation.
Although the observation that yields the strongest gamma ray flux occurs
on the day when the daily averaged RXTE X-ray flux was the highest of
any day on which we observed Cen X-3, there is no evidence for a formal
correlation between the VHE gamma-ray and X-ray fluxes.

We have also tested for modulation of the VHE gamma ray flux at the
orbital period of the binary system and at the pulsar period. We have no
evidence for modulation of the VHE gamma ray emission at either period.
 
The processes considered for the production of TeV gamma rays in X-ray
binaries have included beam dump models with both electrons and protons
as the accelerated particles (Vestrand \& Eichler (1982) for protons and
Cheng, Ho \& Ruderman (1985) for electrons). In addition, a co-rotating
jet model has been suggested by Kiraly \& Meszaros (1988).

We are grateful to the UK Particle Physics and Astronomy Research
Council for support of the project. The Mark 6 telescope was designed
and constructed with the assistance of the staff of the Physics
Department, University of Durham.This paper uses quick look results
provided by the ASM/{\it RXTE} and BATSE teams. Dr. Mark Finger is
thanked for his kind help with BATSE data.


\vspace{1ex}
\begin{center}
{\Large\bf References}
\end{center}
%
Armstrong, P., et al. 1998, Exp. Astro., in press\\
Carraminana, A., et al. 1989, Timing Neutron Stars ed. H. \"{O}gelman \&
E. P. J. van den Heuvel (Dordrecht: Kluwer Academic Press), p~369\\
Chadwick, P. M., et al. 1998, \apj, 503, 391\\
Chadwick, P. M., et al. 1999, \apj, 513, 161\\
Cheng, K. S., Ho, C., \& Ruderman, M. 1985, \apj, 300, 522\\
Fegan, D. J. 1997, J. Phys. G. Nucl. Part. Phys., 23, 1013\\
Hillas, A. M. 1984, Nature, 312, 50\\ 
Hjellming, R. M. \& Han, X. 1995, X-ray Binaries, ed. W. H. G. Lewin, J.
van Paradijs \& E. P. J. van den Heuvel (Cambridge: Cambridge University
Press), p. 308\\
Kelley, R. L., et al. 1983, \apj, 268, 790\\
Kiraly, P., and Meszaros, P. 1988, \apj, 333, 719\\
Raubenheimer, B. C., et al. 1989, \apj, 336, 349\\
Schubnell, M. 1997, Proc. 4th. Compton Symposium, ed. C. D. Dermer, M. S.
Strickman, \& J. D. Kurfess (New York: AIP), 2, 1386\\
Vestrand, W. T., \& Eichler, D. 1979, \apj, 261, 251\\
Vestrand, W. T., Sreekumar, P. \& Mori, M. 1997, \apjl, 483, L49\\
Weekes, T. C., et al. 1997, Proc. 4th. Compton Symposium, ed. C. D.
Dermer, M. S. Strickman, \& J. D. Kurfess (New York: AIP), 1, 361\\
\end{document}